\documentclass[
]{amsproc}

\newtheorem{thm}{Theorem}[section]
\newtheorem{prop}[thm]{Proposition}

\theoremstyle{definition}%
\newtheorem{defn}[thm]{Definition}

\newtheorem{example}[thm]{Example}
\newtheorem{conj}[thm]{Conjecture}

\theoremstyle{remark}

\PassOptionsToPackage{pdfauthor={Vladimir Kisil},%
    pdftitle={Descartes and Klein on Noncommutative Geometry},%
    pdfsubject={Mathematical Physics},%
    backref=page,%
  pdfkeywords={symmetries,Heisenberg group, special linear group,
    symplectic group, Hardy space, Segal-Bargmann space, Clifford
    algebra, Cauchy-Riemann-Dirac operator, Moebius transformations,
    functional calculus, Weyl calculus (quantization), quantum
    mechanics, Schrodinger representation, metaplectic representation}
}{hyperref}

\usepackage{graphicx}
\usepackage[noshort]{myart2k}
\ppnum{}{\href{http://arXiv.org/abs/math-ph/0112059}{math-ph/0112059}}
{2001}

\makeatletter
  \def\p@enumi{\thethm}
  
  \def\theenumi{(\@arabic\c@enumi)}
\makeatother

\input{mydef}

\providecommand{\myhbar}{h}

\providecommand{\matr}[4]{{\ensuremath{ \left(\!\! \begin{array}{cc}
#1 & #2 \\ #3 & #4
\end{array}\!\!\right) }}}

\newcommand{\Ba}{\bar{\alpha}}
\newcommand{\Bb}{\bar{\beta}}

\providecommand{\oper}[1]{\mathcal{#1}}

\newcommand{\n}[1]{\ensuremath{\mathsf{#1}}}

\newcommand{\TSpace}[2]{\ensuremath{ { \widetilde{\mathbb{#1}}^{#2}} }}

\newcommand{\SP}[1][1]{\ensuremath{\FSpace{Sp}{}(#1)}}
\newcommand{\SU}[1][1,1]{\ensuremath{\FSpace{SU}{}(#1)}}
\newcommand{\anti}{\mathcal{A}}
\ifundefined{eoe}
   \DeclareMathSymbol{\eoe}{\mathord}{AMSa}{"06}
\fi
\newcounter{myenumi}
\renewcommand{\themyenumi}{(\alph{myenumi})}
\newenvironment{myenumerate}{\begin{list}{\textbf{\themyenumi}}
    {\usecounter{myenumi}\leftmargin=0pt
      \addtolength{\labelwidth}{-0.5em}
}}
{\hfill$\eoe$\end{list}}

\newcommand{\ub}[3][]{\left\{\!#1\left[#2,#3\right]\!#1\right\}}

\begin{document}

\title[Descartes and Klein on Noncommutative Geometry]{Meeting
  Descartes and Klein\\ Somewhere in a Noncommutative Space}

\author[Vladimir V. Kisil]{\href{http://maths.leeds.ac.uk/~kisilv/}{Vladimir V. Kisil}}

\address{%
School of Mathematics\\
University of Leeds\\
Leeds LS2\,9JT\\
UK
}

\email{\href{mailto:kisilv@maths.leeds.ac.uk}{kisilv@maths.leeds.ac.uk}}

\urladdr{\href{http://maths.leeds.ac.uk/~kisilv/}%
{http://maths.leeds.ac.uk/\~{}kisilv/}}

\begin{abstract}
  We combine the coordinate method and Erlangen program in the
  framework of noncommutative geometry through an investigation of
  symmetries of noncommutative coordinate algebras. As the model we
  use the coherent states construction and the wavelet transform in
  functional spaces. New examples are a three dimensional spectrum of
  a non-normal matrix and a quantisation procedure from
  symplectomorphisms.
\end{abstract}
\thanks{On leave from the Odessa University.}
\maketitle
  \tableofcontents

\section{Introduction}
\label{sec:introduction}

\epigraph{Mathematics is a part of physics. Physics is an experimental
science, a part of natural science. Mathematics is the part of physics
where experiments are cheap.}{Vladimir
Arnold}{\textup{\cite{Arnold98a}}}\medskip  

We used to think by a small number of mental images which help us to
understand equally good (or bad) a variety of different processes. A
``Big Bang'' is one of such pet ideas which we call archetypes to
excuse their overloading.  \person{K.~Jaspers} argued that modern
culture appeared from a black matter as result of a big bang at
the \emph{axial period} more than two thousand years ago. The expanding Universe
of human culture became split into many seemingly independent
galaxies---science, religion, art, etc.---each with its own
complicated structure and dynamics. The cosmological belief that all
future of a Universe was determined by the first minutes after the bang
may not be true but has an appeal of simplicity. Anyway it is natural
to expect that after the bang all parts will run away each other. Thus
the appearance of \emph{two cultures} in the sense of
\person{C.P.~Snow}, which are disjoint or even in an
opposition, seems to be unavoidable.

Yet there is also another persistent pattern: mathematics since
\emph{Elements} of Euclid grown enormously both in qualitative and
quantitative sense but did not split into several smaller independent
subjects. And the border between mathematics and physics (if ever
exists at all) is as thin today as in times of Archimedes. Moreover a
smuggling across that border in both direction is more rewarding today
than ever before. Is there a hidden rules or forces which tie them
together despite of general centrifugal tendencies?

\section{Descartes Meets Klein: Symmetries of Coordinate Algebras}
\label{sec:invetion-coordinates}

\epigraph{If you would be a real seeker after truth, you must at least
  once in your life doubt, as far as possible, all things.} 
{Descartes}{Discours de la M\'ethode, 1637} \medskip

The striking example of centripetal trends in mathematics is the
Cartesian coordinate method. Before the XVII century there were two
big and relatively independent mathematical subjects with different
(even geographically) origins: the \emph{synthetic geometry} of Greeks
and \emph{abstract algebra} of Arabs. It was natural to expect that
these fields will diverge even further during their developments. Thus
the proposition of \person{Descartes} to associate geometrical
problems with algebraic equations by an \emph{introduction of
coordinates} was a great manifestation of the integrity of
mathematics. Another example of unexpected links between seemingly
unrelated topics was the \person{Galois} discovery that solvability of
\emph{algebraic equations} depends on certain \emph{group-theoretic
properties} of their Galois group. Just putting together these two
ideas one may suspect that there is a connection between geometry and
group theory. That connection was announced in the famous Erlangen
program of \person{Felix Klein}
\href{http://www-groups.dcs.st-and.ac.uk/~history/Mathematicians/Klein.html}{developed
under the strong influence of \person{Sophus Lie}}: synthesis of
geometry as the study of the properties of a space that are invariant
under a given group of transformations.

We will not retell once again the story of coordinate approach in
noncommutative spaces, see for example~\cite{Segal96b} for a balanced
and concise exposition. Instead we would highlight few observations
oftenly overlooked in the current literature:
\begin{enumerate}
\item The rule that coordinates should form an \emph{algebra} was
  not introduced by \person{Descartes} originally, it is sufficient
  that coordinates have \emph{any} rich algebraic structure to reflect all
  geometrical properties, for example, via algebraic or differential
  equations. The identification 
  \begin{equation}\label{eq:coord-algebra}
    \mathrm{coordinates}=\mathrm{an associative algebra}+\mathrm{some topology}
  \end{equation} was fixed only after the Gelfand theory of
  commutative Banach algebras. While the achievements based
  on~\eqref{eq:coord-algebra} are really impressive~\cite{Connes00a}
  that identification is not necessary (see bellow) and could be a
  needless \emph{restriction} in general.
\item The development of noncommutative geometry was oftenly
  \emph{motivated} and \emph{supported} by the representation theory
  of groups. For example, the original paper on noncommutative
  measure and integration theory~\cite{Segal53a} leaded to the
  Plancherel theorem for noncommutative locally compact
  groups. Quantum theory---the stronghold of Cartesian approach to
  noncommutative spaces---was able to deal with elementary particles
  or four basic interactions only in the Erlangen spirit through
  their symmetry groups. 
\item The Cartesian connection of geometrical problems with algebraic
  equations was not a subordination of geometry to algebra, actually
  Descartes used it in both ways and introduced a
  \href{http://www-groups.dcs.st-and.ac.uk/~history/Mathematicians/Descartes.html}{geometrical
  method for solutions of quadratic equations}. Forever geometry be
  seen as an extremely beautiful subject with its own charm: ``a
  geometrical proofs'' usually means ``an elegant proof'', and its is
  fashionable to say ``I am doing noncommutative geometry'' rather
  than ``I am studying operator algebras and applications''.
\end{enumerate}

We challenge the rigid identification~\eqref{eq:coord-algebra} in this
paper and use the assumption:
\begin{equation}\label{eq:coord-homogen}
  \textit{coordinates are oftenly a representation space for a group action}
\end{equation} to approach certain problem in noncommutative spaces.
The fact that sometimes coordinates form also an algebra \emph{could
be} useful but \emph{is not} crucial anymore. We will see in the last
two sections that it is even helpful to downplay the structure of
algebras by abandoning the algebraic homomorphism property. The number
of applications is not limited to given in the present paper, (see in
addition Example~2 in~\cite{Kisil97a} with the Manin plane and quantum
groups~\cite{Manin91}) and they deserve a further investigation.

One can object~\cite{Connes00a} that homogeneous spaces which are
geometries in the sense of the Klein program are too restrictive to
give a good model of space-time in general relativity. There is no a
hard evidence to refute that claim at the moment, but we could learn
from the inspiring paper~\cite{Cartier01a} that symmetries of objects
are usually richer than people ordinary think. For example, even the
structure of a single point could be significantly enriched by
introduction of a coordinate bundle over it~\cite[\S~6]{Cartier01a}
and assigning a group action in that bundle. Therefore one could make
the following conjecture, which we illustrate by examples in the
present paper.

\begin{conj}
  The combination of Cartesian and Klein-Lie approaches based on the
  assumption~\eqref{eq:coord-homogen} is 
  stronger than the original Erlangen program itself and could go
  beyond previous limits.
\end{conj} 

The rest of the paper is organised as follows. In the next Section we
will study symmetries of functional spaces which are coordinates in
commutative cases. This will be our platform for an invasion to
noncommutative spaces, similarly to the Gelfand structural theorem
about commutative Banach algebras in the approach based on
identification~\eqref{eq:coord-algebra}. The technique is widely
known as coherent states construction and wavelet transform but was
rediscovered many times before and after those names were coined. We
show that many fundamental notions of analysis are intimately
connected to that circle of ideas, which are yet not explicitly
understood and used to its full power. 

Intertwining commutative and noncommutative coordinates in
Section~\ref{sec:funct-calc-as} we will get a new description of
functional calculus and related spectrum of non-normal matrices.  The
calculus and the spectrum are naturally connected through an
appropriately extended spectral mapping theorem.

We also approach the quantisation problem with the similar ideas in
Section~\ref{sec:quant-from-sypl} and get a natural combination of
quantum and classical mechanics within the framework of the Heisenberg
group.

This paper is a survey or even an essay on the subject. The three
main sections are rather connected by a common idea than technically
dependent. Therefore they could be looked through almost separately. More
details could be found in published
papers~\cite{Kisil95i,Kisil97c,Kisil98a,Kisil00a} and 
will also appear in~\cite{Kisil02b,Kisil02a}.

\section{Coherent States and Wavelets in Mathematics and Physics}
\label{sec:coher-stat-wavel}

\epigraph{In the 1960's it was said (in a certain connection) that the
  most important discovery of recent years in physics was the complex
  numbers.}{Yu.I.~Manin}{\textup{\cite[Preface]{Manin81a}}} \medskip

We would like to present a construction which produces many important
objects in analytic function theory, i.e. commutative coordinate
spaces, out of symmetry groups.  The scheme is well known,
cf.~\cite{AliAntGaz,Perelomov86} and got much attention in recent
decades but it is not used to its full potential yet.  Our main
examples are provided by the one dimensional Heisenberg group
$\Space{H}{}=\Space{H}{1}$~\cite{Folland89,MTaylor86} and the $\SL$
groups~\cite{HoweTan92,Lang85,MTaylor86}---two groups with the outmost
importance~\cite{Howe80a,HoweTan92} in both mathematics and physics.

The one dimensional Heisenberg group $\Space{H}{}$ consists of
points $(s,z)=(s,x,y)$ parametrised by $s\in\Space{R}{}$ and
$z=x+iy\in \Space{C}{}$. The group law is given by:
\begin{equation}\label{eq:g-heisenberg}
  g*g'=(s,z)*(s',z')=(s+s'+\frac{1}{2}\Im(\bar{z}z'), z+z'),
\end{equation} where $\Im w$ denotes the imaginary part of a complex
number $w$.  The $\Space{H}{}$ is the necessary component (sometimes
implicit) of any quantisation scheme because its Lie algebra has the
only non-trivial commutator:
\begin{displaymath}[]
  [X,Y]=S,
\end{displaymath} which in the Shr\"odinger representation (see bellow
\eqref{eq:schrodinger}) takes a form of the celebrated Heisenberg
\emph{uncertainty relation}.

The group $\SL$ consists of $ 2\times 2$ matrices $
\matr{a}{b}{c}{d}$ with real entries and determinant $ad-bc=1$. Its is
also isomorphic to the following three groups \cite[\S~8.1]{MTaylor86}
which wee will use bellow in different contexts:
\begin{enumerate}
\item the Lorentz type group $SO_e(1,1)$ in the two-dimensional Minkowski
  space with the metric $ds^2=dt^2-dx^2$;
\item the group $\SU$ of linear transformation of $\Space{C}{2}$
  preserving the quadratic form $z_1^2-z_2^2$;
\item the symplectic group $\SP$ of linear symplectomorphisms of the
  two dimensional flat phase space in classical mechanics.
\end{enumerate}

It is not surprising that $\Space{H}{}$ and $\SL$ are intimately
connected to each other as we will see and use in the
Section~\ref{sec:quant-from-sypl}.

\subsection{Space-Time or Phase Space from Symmetry Groups}

There is a heuristic observation~\cite{Segal96b} that a
\emph{space-time} is not a primary concept but appears in the approach
based on the identification~\eqref{eq:coord-algebra} as the spectrum
of a maximal commutative subalgebra of the algebra of observables
invariant under the fundamental group.  We show in this subsection
that simplest forms of the space-time and the \emph{phase space} could
be naturally obtained just from symmetry groups in the context
of~\eqref{eq:coord-homogen}. It is interesting to note that the same
group could produce several rather distinct spaces,
e.g. with elliptic or hyperbolic metrics. 

Abstract scheme could be described as follows.  Let $G$ be a group and
$H$ be its closed normal subgroup, which could be trivially just
$\{e\}$.  Let $X=G/H$ be the corresponding homogeneous space with an
invariant measure $d\mu$ and $s: X \rightarrow G$ be a Borel section
in the principal bundle $G \rightarrow
G/H$~\cite[\S~13.2]{Kirillov76}.  Then any $g\in G$ has a unique
decomposition of the form $g=s(x)h$, $ x\in X$ and we will write $
x=s^{-1}(g)$, $h=r(g)={(s^{-1}(g))}^{-1}g$.  Note that $ X $ is a left
$G$-homogeneous space with an action defined in terms of $s$ as
follow: 
\begin{equation}\label{eq:action-on-X}
  g: x \mapsto g\cdot x=s^{-1}(g^{-1}* s(x))
\end{equation}
where $*$ is
the multiplication on $G$. 

We will illustrate our consideration by a chain of examples. Each one
consists of four parts numbered from (a) to (d): two cases for
$G=\Space{H}{}$ with two different subgroups $H=\Space{R}{2}$ and
$H=\Space{R}{}$, see~\cite{Kisil98a} for more details; other two cases
for $G=\SL$ with subgroup $H=K$ and $H=A$ studied in~\cite{Kisil97c}.
\begin{example}
\label{ex:G-acts-on-X}
\begin{myenumerate}
\item \label{it:H-on-R}
  We start from $\Space{H}{}$ and its subgroup
  $H=\Space{R}{2}=\{(t,z)\such \Im(z)=0\}$.  Then $X=G/H=\Space{R}{}$
  and because the Haar measure on $ \Space{H}{} $ coincides with the
  standard Lebesgue measure on $\Space{R}{3}$~\cite[\S~1.1]{MTaylor86}
  then invariant measure on $X$ coincides with the Lebesgue measure on
  $\Space{R}{}$.
  Mappings $s: \Space{R}{} \rightarrow \Space{H}{}$ and $r:
  \Space{H}{} \rightarrow H$ are defined by the identities
  $s(x)=(0,ix)$, $s^{-1}(t,z)=\Im z$, $r(t,u+iv)=(t,u)$.  The
  composition law $s^{-1}((t,z)\cdot s(x))=x+u$ reduces to Euclidean
  shifts on $\Space{R}{}$.  We also find $s^{-1}((s(x_1))^{-1}\cdot
  s(x_2))=x_2-x_1$ and $r((s(x_1))^{-1}\cdot s(x_2))= 0$. This $X$ is
  the \emph{configuration space} of a particle with one degree of
  freedom. 
  
\item \label{it:H-on-C}
  As a subgroup $H=\Space{R}{}$ we select now the one dimensional
  centre of $\Space{H}{}$ consisting of elements $(s,0)$.  Of course
  $X=G/H$ isomorphic to $\Space{C}{}$ and mapping $s: \Space{C}{}
  \rightarrow G$ simply is defined as $s(z)=(0,z)$.   The invariant measure
  on $X$ also coincides with the Lebesgue measure on $\Space{C}{}$.
  Note also that composition law $s^{-1}(g\cdot s(z))$ reduces to
  Euclidean shifts on $\Space{C}{}$.  We also find
  $s^{-1}((s(z_1))^{-1}\cdot s(z_2))=z_2-z_1$ and
  $r((s(z_1))^{-1}\cdot s(z_2))= \frac{1}{2} \Im \bar{z}_1z_2$--the
  \emph{symplectic form} on $\Space{R}{2}$. In that case we get the
  \emph{phase space} of a particle with one degree of freedom.

  \begin{figure}[tbh]
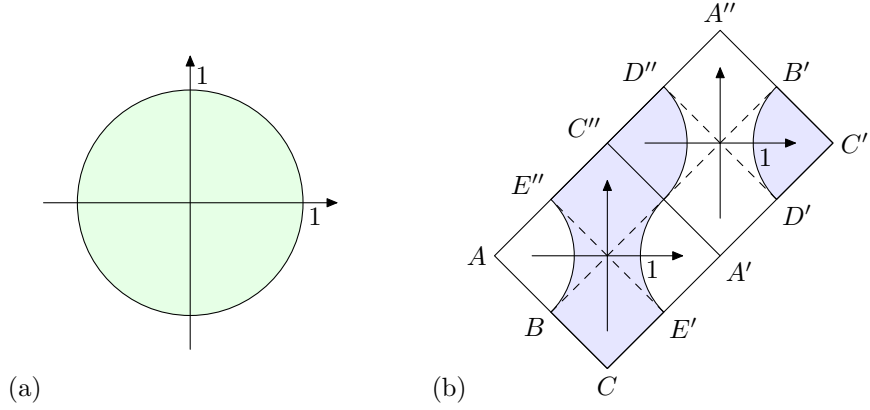

    \begin{center}
      \href{http://maths.leeds.ac.uk/~kisilv/r11.gif}{
      (a)\includegraphics{nccg2.1} \hspace{1cm}
      (b)\includegraphics{nccg2.2}}
      \caption[Two unit disks]
      {
        Two unit disks in elliptic (a) and hyperbolic (b)
        metrics. In (b) squares $ACA'C''$ and $A'C'A''C''$ represent
        two copies of $\Space{R}{2}$, their boundaries are the image
        of the light cone at infinity. These cones should be glued in
        a way to merge points with the same letters (regardless number
        of dashes). The hyperbolic unit disk $\TSpace{D}{}$ (shaded
        area) is bounded by four branches of hyperbola. Dashed lines
        are light cones at origins.}
      \label{fig:unit-disk}
    \end{center}
  \end{figure}
  
\item \label{it:SL-ellipt}
  Here we study $ \SL $  in the form of the group $SU(1,1)$ of $
  2\times 2$ matrices with complex entries of the form $ \matr{ \alpha}{
    \beta}{ \bar{\beta}}{\bar{ \alpha}}$ such that $ \modulus{ \alpha }^2 - 
  \modulus{ \beta }^2=1 $. 
  $ \SL $ has the only non-trivial compact closed subgroup
  $K$, namely the group of matrices of the form
  $h_{\psi}=\matr{e^{i\psi}}{0}{0}{e^{-i\psi}}$.
  Any $ g\in \SL $ has a unique decomposition of the
  form 
  \begin{equation}\label{eq:sl2-compl-decomp}
    \matr{\alpha}{\beta}{\bar{\beta}}{\bar{\alpha}} 
     =  \frac{1}{\sqrt{1- \modulus{a}^2 }}
    \matr{1}{a}{\bar{a}}{1} 
    \matr{e^{i\psi}}{0}{0}{e^{-i\psi}} 
  \end{equation}
  where $\psi=\Im \ln \alpha$, $a=\beta\bar{\alpha}^{-1}$, and $
  \modulus{ a } < 1$ because $ \modulus{ \alpha }^2 - \modulus{ \beta
    }^2=1 $. Thus we can identify $\SL / H$ with the unit
  disk $ \Space{D}{} $, see Figure~\ref{fig:unit-disk}(a), and define
  mapping $s: \Space{D}{} \rightarrow \SL $ and $ r: G \rightarrow H$
  as follows
  \begin{equation} \label{eq:def-s-a}
    s: a \mapsto \frac{1}{\sqrt{1- \modulus{a}^2 }}
    \matr{1}{a}{\bar{a}}{1}\qquad \qquad
    r: \matr{\alpha}{\beta}{\bar{\beta}}{\bar{\alpha}} \mapsto \matr{ 
      \frac{{\alpha}}{
        \modulus{\alpha} }}{0}{0}{\frac{\bar{\alpha}}{ \modulus{\alpha} }}.
  \end{equation}

  The formula $g: a \mapsto g\cdot a=s^{-1}(g^{-1} * s(a)) $
  associates with a matrix $ g^{-1}=\matr{ \alpha}{ \beta}{
  \bar{\beta}}{ \bar{\alpha}}$ the fraction-linear \emph{conformal}
  transformation of $ \Space{D}{} $ of the form
  \begin{equation} \label{eq:fr-lin-a}
    g: z \mapsto g\cdot z=
    \frac{\alpha z + \beta}{\bar{ \beta}z +\bar{\alpha}},  
    \qquad g^{-1}=\matr{\alpha}{\beta}{\bar{\beta}}{\bar{\alpha}},
  \end{equation}
  which also can be extended to a transformation of $\dot{ \Space{C}{} }$
  (the one-point compactification of $ \Space{C}{} $). Here
  $X=\Space{D}{}$ is two dimensional \emph{conformal configuration space}.

\item \label{it:SL-hyperbolic}
  Now we use $\SL$ in the form of a Lorentz type group $SO_e(1,1)$. It
  is convenient to represent its elements again as $2\times
  2$-matrices but this time with Clifford algebra values. This four
  real dimensional Clifford algebra
  $\Cliff{1,1}$~\cite{DelSomSou92,Porteous95} generated by $1$ and two
  imaginary units $e_1$ and $e_2$ such that
  \begin{displaymath}
    e_1^2=-e^2_2=-1, \qquad e_1e_2=-e_2e_1.
  \end{displaymath}
  We use \textsf{Sanserif font} for elements of
  $\Cliff{1,1}$. Then $\SL$ is represented~\cite{Kisil97c} by matrices
  $\matr{\n{a}}{\n{b}}{-\n{b}}{\n{a}}$ where
  $\n{a}\bar{\n{a}}-\n{b} \bar{\n{b}}=1$. 
  We have a decomposition similar to \eqref{eq:sl2-compl-decomp}:
  \begin{equation}
    \matr{\n{a}}{\n{b}}{-\n{b}}{\n{a}} = 
    \modulus{\n{a}} \matr{1}{\n{b}\n{a}^{-1}}{-\n{b}\n{a}^{-1}}{1}   \matr{ 
      \frac{\n{a}}{ \modulus{\n{a}}} }{0}{0}{\frac{\n{a}}{ \modulus{\n{a}}}}.
  \end{equation}
  It could be seen~\cite{Kisil97c} that $\n{b}\n{a}^{-1}\in
  \Space{R}{1,1}$, i.e. is a two-dimensional vector in Minkowski
  space.  But now we could \emph{not} get that
  $\modulus{\n{b}\n{a}^{-1}} <1 $, or equivalently we could not separate:
  \begin{enumerate}
  \item \emph{topologically} the Minkowski space $\Space{R}{1,1}$ into
    \emph{interior} and \emph{exterior} of the unit circle; 
  \item \emph{analytically} $\FSpace{L}{2}$ on the unit ``circle'' into
    subspaces of \emph{analytic} and \emph{antianalytic} 
    functions; 
  \item \emph{physically} the time axis into the \emph{future}
    and the \emph{past} halves,
  \end{enumerate} because
  \href{http://maths.leeds.ac.uk/~kisilv/r11.gif}{M\"obius
  (linear-fractional) transformations mix both sets in each case}. A
  way out is known and is the same both in physical~\cite{Segal76} and
  mathematical~\cite{Kisil97c} situations: we need to take a double
  cover of $\Space{R}{1,1}$ and chose the ``unit disk'' $\TSpace{D}{}$
  as shown on Figure~\ref{fig:unit-disk}(b) and explained in its
  caption.

  Matrices of the form 
  $
  \matr{\n{a}}{0}{0}{\n{a}}=\matr{e^{e_1e_2\tau}}{0}{0}{e^{e_1e_2\tau}}$, 
  $\n{a}={e^{e_1e_2\tau}}=\cosh\tau+e_1e_2\sinh\tau, \quad \tau\in 
  \Space{R}{}
  $ 
  comprise a subgroup of hyperbolic rotations in $\Space{R}{1,1}$
  which we denote by $A$. 
  We define an embedding $s$ of $\TSpace{D}{}$ for our realization of 
  $\SL$ by the formula:
  \begin{equation} \label{eq:def-s-b}
    s: \n{u} \mapsto \frac{1}{ \sqrt[]{1+\n{u}^2} }
    \matr{1}{\n{u}}{-\n{u}}{1}, \qquad
    r: \matr{\n{a}}{\n{b}}{-\n{b}}{\n{a}} \mapsto \matr{ \frac{\n{a}}{ 
        \modulus{\n{a}} }}{0}{0}{\frac{\n{a}}{ \modulus{\n{a}} }}
  \end{equation}
  The formula $g: \n{u} \mapsto s^{-1}(g\cdot s(\n{u}))$ 
  gives the linear-fraction transformation $ \TSpace{D}{} \rightarrow
  \TSpace{D}{}$ conformal in the hyperbolic metric:
  \begin{equation} \label{eq:fr-lin-b}
    g: \n{u} \mapsto g\cdot \n{u}= \frac{\n{a}\n{u}+\n{b}}{-\n{b}\n{u}+\n{a}}, 
    \qquad g^{-1}=\matr{\n{a}}{\n{b}}{-\n{b}}{\n{a}},
  \end{equation} which is similar to~\eqref{eq:fr-lin-a}. We get
  two dimensional \emph{relativistic space-time}. The appearance of
  Clifford algebra in relativistic case is
  expectable~\cite{DelSomSou92,Porteous95}.  
\end{myenumerate}
\end{example}
In the following we call $X$ just \emph{space} understanding that in
different realisations it could be either a configuration space or
phase space or space-time.

\subsection{The Vacuum and Reduced Wavelet Transform} 

We are ready to explain the r\^ole of the subgroup $H$ in the coherent
states construction: it selects the \emph{vacuum} as its eigenvector
among all possible physical states. Thereafter having the chosen
vacuum we can generate all possible states of the system from a
\emph{representation} $\rho$ of $G$ by isometric operators in a Banach
space $B$. Coherent states are parametrised by points of the space
$X=G/H$.

\begin{defn} \label{de:coherent1}
  \textup{\cite{Kisil98a}}
  Let $G$, $H$, $X=G/H$, $s: X \rightarrow G$, $\rho: G \rightarrow
  \FSpace{L}{}(B)$ be as before. We say that $b_0 \in
  B$ is a \emph{vacuum 
    vector} if for all $h\in H$
  \begin{equation} \label{eq:h-char}
    \rho(h) b_0 = \chi(h) b_0, \qquad \chi(h) \in \Space{C}{}.
  \end{equation} We say that set of vectors $b_x=\rho(x) b_0$, $x\in
  X$ form a family of \emph{coherent states} if there exists a
  continuous non-zero linear functional $l_0 \in B^*$, called the
  \emph{analysing functional}, such that
  \begin{enumerate}
  \item \label{it:norm} $\norm{b_0}=1$, $\norm{l_0}=1$,
    $\scalar{b_0}{l_0}\neq 0$;
  \item \label{it:h-char} $\rho(h)^* l_0=\bar{\chi}(h) l_0$, where
    $\rho(h)^*$ is the adjoint operator to $\rho(h)$;
  \item \label{it:coher-eq} The following equality holds
    \begin{equation} \label{eq:coher-eq}
      \int_X \scalar{\rho(x^{-1}) b_0}{l_0}\, \scalar{\rho(x)
      b_0}{l_0}\, d\mu(x) = \scalar{b_0}{l_0}.
    \end{equation}
  \end{enumerate}
\end{defn}

With analysing functional we are able to decompose any state as a
superposition of the coherent states. 

\begin{defn}
  The \emph{reduced wavelet transform} $\oper{W}$ from a Banach space
  $B$ to a space of function $\FSpace{F}{}(X)$ on a homogeneous space
  $X=G/H$ defined by a representation $\rho$ of $G$ on $B$, a vacuum
  vector $b_0$ and a test functional $l_0$ is:
  \begin{equation} \label{eq:wave-tr}
    \oper{W}: v \mapsto \hat{v}(x)= 
    [\oper{W}v] (x)=\scalar{\rho(x^{-1}) v}{l_0}=
    \scalar{v}{\rho^*(x)l_0}.
  \end{equation}
  The \emph{inverse wavelet transform} $\oper{M}$ from $
  \FSpace{F}{}(X) $ to $B$ is given by the formula:
  \begin{equation}\label{eq:m-tr}
    \oper{M}:  \hat{v}(x) \mapsto \oper{M} 
    [\hat{v}(x)]  =  \int_X \hat{v}(x) b_x\,d\mu(x) 
     = \int_X \hat{v}(x) \rho(x)\,d\mu(x) b_0.
  \end{equation}
\end{defn}

The geometric action~\eqref{eq:action-on-X} of $G: X \rightarrow X$
defines a representation $\lambda(g): \FSpace{F}{}(X) \rightarrow
\FSpace{F}{}(X)$ induced by a character $\chi$ of $H$ as follows
\begin{equation} \label{eq:l-rep}
[\lambda(g) f] (x) = \chi(r(g^{-1}\cdot x)) f(g^{-1}\cdot x).
\end{equation}
For the case of trivial $H=\{e\}$ the representation~\eqref{eq:l-rep}
becomes the left regular representation $\rho_l(g)$ of $G$ on
$\FSpace{L}{2}(G)$. 
\begin{prop} \label{pr:inter1}
  We have:
  \begin{enumerate}
  \item The reduced wavelet transform $\oper{W}$  and the inverse
    wavelet transform $ \oper{M} $ intertwine $\rho$ and the
    representation $\lambda$~\eqref{eq:l-rep} on $\FSpace{F}{}(X)$:
  \begin{equation}\label{eq:W-M-intertwine}
    \oper{W} \rho(g) = \lambda(g) \oper{W} \qquad \mathrm{and} \qquad 
    \oper{M} \lambda(g) = \rho(g) \oper{M} \qquad 
    \mathrm{ for all } g\in G.
  \end{equation}
  \item There is an isomorphism property:
    \begin{equation} \label{eq:isom1}
      \scalar{\oper{W} v }{ \oper{M}^* l}_{ \FSpace{F}{}(X) } =
      \scalar{v}{l}_B,  
      \qquad \forall v\in B, \quad l\in B^*
    \end{equation}
  \item The image $\FSpace{F}{}(X)$ of $B$ under $\oper{W}$ is
    $\lambda$-invariant subspace of $\FSpace{C}{}(G)$.
  \end{enumerate}
\end{prop}

There is a physical meaning of~\eqref{eq:W-M-intertwine}: having a
representation $\rho$ of $G$ in an abstract space $\FSpace{L}{}(B)$ of
\emph{observables} we could introduce a space $X$ and realise $B$ as
functions on $X$ with a geometric action $\lambda$~\eqref{eq:l-rep}
instead of an abstract $\rho$. This advantage allows us to use the
\emph{intertwining property}~\eqref{eq:W-M-intertwine} in new
Definition~\ref{de:func-calc-sl2} of functional calculus and
Definition~\ref{de:quantisation} of quantisation instead of
the traditional \emph{algebraic homomorphism} property. 

\begin{thm}
  The composition of transforms $\oper{M}$ and $\oper{W}$
  \begin{equation} \label{eq:szego}
    \oper{P}= \oper{M} \oper{W}: B \rightarrow B
  \end{equation}
  is a projection of $B$ to its linear subspace for which $b_0$ is
  cyclic. Particularly if $\rho$ is an irreducible representation then the
  inverse wavelet transform $\oper{M}$ is a \emph{left inverse} operator
  on $B$ for the wavelet transform $\oper{W}$:
  $
    \oper{M}\oper{W}=I.
  $
\end{thm}

\begin{example}
  \label{ex:wavelets}
\begin{myenumerate}
\item 
  We take a representation $\sigma_\myhbar$ of $\Space{H}{}$ in
  $\FSpace{L}{p}(\Space{R}{})$, $1<p<\infty$ by operators of shift and
  multiplication~\cite[\S~1.1]{MTaylor86}:
  \begin{equation} \label{eq:schrodinger}
    \sigma_\myhbar(s,z): f(y) \rightarrow [\sigma_\myhbar(s,z)f](y)
    =e^{i(2s\myhbar-\sqrt{2\myhbar} vy+ \myhbar uv)} 
    f(y- \sqrt{2\myhbar}u), \qquad z=u+iv,
  \end{equation} It is the \emph{Schr\"odinger representation} with
  parameter $\myhbar$.  As a character of $H=\Space{R}{2}$ we take the
  $\chi(s,u)=e^{2it}$.  The corresponding test functional $l_0$
  satisfying to~~\ref{it:h-char} is the integration
  $l_0(f)=(2\pi)^{-n/2}\int_{\Space{R}{n}} f(y)\,dy$.  Thus the
  wavelet transform~~\eqref{eq:wave-tr} is
  \begin{equation}\label{eq:fourier}
    \hat{f}(x)=\int_{\Space{R}{n}} \sigma(s(x)^{-1}) f(y)\,dy =
    (2\pi)^{-n/2}\int_{\Space{R}{n}} e^{i \sqrt{2}xy} f(y)\,dy
  \end{equation} and is nothing else but the Fourier transform.
  There is no a vacuum vector in our space $B$, see physical
  implications of an absence of vacuum in~\cite{Segal90}. We however
  could proceed as in~\cite[\S~2.3]{Kisil98a} and take a bigger space
  $B'=\FSpace{L}{\infty}(\Space{R}{n})\supset B$ with the vacuum
  vector $b_0(y)\equiv (2\pi)^{-n/2} \in B'$. Then coherent states are
  $b_x(y)=(2\pi)^{-n/2} e^{-i \sqrt{2}xy}$ and the inverse wavelet
  transform~\eqref{eq:m-tr} is defined by the inverse Fourier transform
  \begin{displaymath}
    f(y)= \int_{\Space{R}{n}} \hat{f}(y) b_x(y)\,dx
    = (2\pi)^{-n/2} \int_{\Space{R}{n}} \hat{f}(y) e^{-i \sqrt{2}xy}\,dx.
  \end{displaymath}
  $\oper{W}$ and $\oper{M}$ intertwine the
  left regular representation --- multiplication by $e^{i \sqrt[]{2}yv}$
  with operators
  \begin{eqnarray*}                          
    [\lambda(g) f] (x) &=& \chi(r(g^{-1}\cdot x)) f(g^{-1}\cdot x)
    = e^{i \sqrt[]{2}\cdot 0} f(x-\sqrt[]{2}u)=f(x-\sqrt[]{2}u),
  \end{eqnarray*}
  i.e. with Euclidean shifts.  From~\eqref{eq:isom1} follows the
  Plancherel identity: 
  \begin{eqnarray*}
    \int_{\Space{R}{n}} \hat{v}(y) \hat{l}(y)\, dy &=&
    \int_{\Space{R}{n}} {v}(x) {l}(x)\, dx .
  \end{eqnarray*}
  Thus our construction generate all important properties of the
  Fourier transform. 

  The Schr\"odinger representation is irreducible on
  $\mathcal{S}(\Space{R}{n})$ thus $\oper{M}=\oper{W}^{-1}$.
  Thereafter the operator~\eqref{eq:szego} representing operators
  $\oper{M}\oper{W}=\oper{W}\oper{M}=1$ correspondingly give an
  integral resolution of the Dirac delta $\delta(x)$:
  \begin{displaymath}[]
    \delta(x-y)=(2\pi)^{-n/2}\int_{\Space{R}{n}} e^{i\xi(x-y)}\,d\xi.
  \end{displaymath}

\item
  As a subgroup $H$ we select now the center of $\Space{H}{}$
  consisting of elements $(s,0)$.  
  As a ``vacuum vector'' we will select the original \emph{vacuum
    vector} of quantum mechanics---the Gauss function
  $f_0(x)=e^{-x^2/2}$ which belongs to all
  $\FSpace{L}{p}(\Space{R}{n})$.  Its transformations are as
  follow:
  \begin{eqnarray*}
    f_g(x)=[\rho_{(s,z)} f_0](x) & = & e^{2it-z\bar{z}/2}e^{-
      (\bar{z}^2+x^2)/2+\sqrt{2}\bar{z}x}. 
  \end{eqnarray*} Particularly $[\rho_{(s,0)} f_0](x)=e^{-2it}f_0(x)$,
  i.e., it really is a vacuum vector in the sense of our
  Definition~\ref{de:coherent1} with respect to $H$.  For the same
  reasons we could take $l_0(x)=e^{-x^2/2} \in
  \FSpace{L}{q}(\Space{R}{n})$, $p^{-1}+q^{-1}=1$ as the analysing
  functional.
  
  Coherent states $f_g(x)=[\rho_{(0,z)} f_0](x)$ belongs to
  $\FSpace{L}{q}(\Space{R}{n})\otimes \FSpace{L}{p}(\Space{C}{n})$ for
  all $p>1$ and $q>1$, $p^{-1}+q^{-1}=1$.  Thus
  transformation~\eqref{eq:wave-tr} with the kernel $[\rho_{(0,z)}
  f_0](x)$ is an embedding $\FSpace{L}{p}(\Space{R}{n}) \rightarrow
  \FSpace{L}{p}(\Space{C}{n})$ and is given by the formula
  \begin{eqnarray}
    \hat{f}(z)&=&e^{-z\bar{z}/2}\pi^{-n/4}\int_{\Space{R}{n}} f(x)\,e^{-
      (z^2+x^2)/2+\sqrt{2}zx}\,dx \label{eq:tr-bargmann}.
  \end{eqnarray}
  Then $\hat{f}(g)$ belongs to $\FSpace{L}{p}( \Space{C}{n} , dg)$
  or its preferably to say that function
  $\breve{f}(z)=e^{z\bar{z}/2}\hat{f}(t_0,z)$ belongs to space
  $\FSpace{L}{p}( \Space{C}{n} , e^{- \modulus{z}^2 }dg)$ because
  $\breve{f}(z)$ is analytic in $z$.  Such functions for $p=2$ form the
  \emph{Segal-Bargmann space} $ \FSpace{F}{2}( \Space{C}{n}, e^{-
    \modulus{z}^2 }dg) $ of functions~\cite{Bargmann61,Segal60}, which
  are analytic by $z$ and square-integrable with respect the Gaussian
  measure $e^{- \modulus{z}^2}dz$.  

  The integral in~\eqref{eq:tr-bargmann} is the well-known
  Segal-Bargmann transform~\cite{Bargmann61,Segal60}. The inverse is
  given by a realization of~\eqref{eq:m-tr}:
  \begin{equation}\label{eq:sb-inverse}
    f(x) = \int_{
      \Space{C}{n} } \breve{f}(z) e^{-
      (\bar{z}^2+x^2)/2+\sqrt{2}\bar{z}x}\, e^{- \modulus{z}^2}\, dz.
  \end{equation} The corresponding operator
  $\oper{P}$~\eqref{eq:szego} is an identity operator $
  \FSpace{L}{p}(\Space{R}{n}) \rightarrow \FSpace{L}{p}(\Space{R}{n})
  $ and~\eqref{eq:szego} gives another integral presentation of the
  Dirac delta.
  
  Integral transformations~\eqref{eq:tr-bargmann}
  and~\eqref{eq:sb-inverse} intertwines the Schr\"odinger
  representation~\eqref{eq:schrodinger} with the following realization
  of representation~\eqref{eq:l-rep}:
  \begin{eqnarray}
    \lambda(s,z) f(w) & = & \hat{f}_0(z^{-1}\cdot w) 
    \bar{\chi}(t+r(z^{-1}\cdot w))
    =  \hat{f}_0(w-z)e^{it+i\Im(\bar{z}w)} 
  \end{eqnarray}
  
  Meanwhile the orthoprojection $\FSpace{L}{2}( \Space{C}{n}, e^{-
    \modulus{z}^2 }dg) \rightarrow \FSpace{F}{2}( \Space{C}{n}, e^{-
    \modulus{z}^2 }dg)$ is of a separate interest and is a principal
  ingredient in Berezin quantization~\cite{Berezin88,Coburn94a}.  Its
  integral kernel is 
  \begin{eqnarray*}
    K(z,w) & = &\exp\left(\frac{1}{2}(- \modulus{z}^2- \modulus{w}^2) 
      +w\bar{z}\right).
  \end{eqnarray*}
  
\item
  We continue with the case of $G=\SL$ and $H=K$. The compact group
  $K\sim \Space{T}{}$ has a discrete set of characters $ \chi_m(h_\phi)=
  e^{-i m\phi} $, $m\in \Space{Z}{} $. We consider here only  $\chi_1$,
  see~\cite{Kisil97c} for others.
  Let us take $X=\Space{T}{}$---the unit circle 
  equipped with the standard Lebesgue measure $ d\phi $ normalised in such a
  way that 
  \begin{equation} \label{eq:n-measure}
    \int_{ \Space{T}{} } \modulus{f_0(\phi)}^2\, d\phi=1 \mathrm{ with }
    f_0(\phi)\equiv 1. 
  \end{equation}
  From ~\eqref{eq:def-s-a} one can find that
  \begin{displaymath}
    r(g^{-1}*s(e^{i\phi}))= \frac{ \bar{\beta}e^{i\phi}+\bar{\alpha}}
    { \modulus{ \bar{\beta}e^{i\phi}+\bar{\alpha}} }, 
    \qquad
    g^{-1}=\matr{\alpha}{\beta}{ \bar{\beta} }{ \bar{\alpha} }.
  \end{displaymath}
  Then the action of $G$ on $\Space{T}{}$ defined by~\eqref{eq:fr-lin-a}, 
  the equality  $ {d(g\cdot \phi)}/{d\phi}= \modulus{\bar{\beta} e^{i\phi} + 
    \bar{\alpha} }^{-2} $  and the character $ \chi_1 $ give the following
  realization of the formula~\eqref{eq:l-rep}: 
  \begin{equation} \label{eq:g-transform}
    [\rho_1(g) f](e^{i\phi})= \frac{1}{ \bar{\beta} e^{i\phi} + \bar{ \alpha }} 
    f \left( \frac{  { \alpha }e^{i\phi}+{\beta}}{\bar{\beta} e^{i\phi} 
        + \bar{ \alpha }} \right).
  \end{equation} This is a unitary representation---the \emph{mock
  discrete series} of $ \SL$~\cite[\S~8.4]{MTaylor86}. It is easily
  seen that $K$ acts in a trivial way~\eqref{eq:h-char} by
  multiplication by $\chi(e^{i\phi})$.  The function $
  f_0(e^{i\phi})\equiv 1 $ mentioned in~\eqref{eq:n-measure}
  transforms
  $
  [\rho_1(g) f_0](e^{i\phi})= (\bar{\beta} e^{i\phi} + \bar{ \alpha 
    })^{-1}
  $ 
  and in particular has an obvious property
  $[\rho_1(h_\psi)f_0](\phi)=e^{i\psi} f_0(\phi)$, i.e.  it is a \emph{vacuum
    vector} with respect to the subgroup $H$.  The smallest linear subspace $
  \FSpace{F}{2}(X) \in \FSpace{L}{2}(X) $ spanned by all these
  transformations consists of boundary values of analytic functions in
  the unit disk and is the \emph{Hardy space}.  Now the reduced
  wavelet transform~\eqref{eq:wave-tr} takes the form~\cite{Kisil97c}:
  \begin{eqnarray}
    \hat{f}(a)=[\oper{W} f] (a) & =  &
    \frac{\sqrt{ 1-\modulus{a}^2 }}{i}
    \int_{\Space{T}{}}  \frac{f(z)}{{a}+z}\,dz, \label{eq:cauchy}
  \end{eqnarray}
  where $z=e^{i\phi}$. Of course~\eqref{eq:cauchy} is the \emph{Cauchy
    integral formula} up to factor ${2\rho } {\sqrt{ 1-\modulus{a}^2
      }} $. The inverse wavelet transform $\oper{M}$ gives an integral
  expression for \emph{orthoprojection Szeg\"o} onto the Hardy space.
  
\item
  Now we consider the same group $G=\SL$ but pick up another subgroup
  $H=A$. Let $e_{12}$ denote $e_1e_2$. The mapping from the subgroup $A\sim
  \Space{R}{}$ to even Clifford numbers $\chi_\sigma: a \mapsto
  a^{e_{12}\sigma}= {(\exp (e_1e_2 \sigma\ln{a}))}= (a \n{p}_1 +
  a^{-1} \n{p}_2) ^\sigma$ parametrised by $\sigma \in \Space{R}{}$ is
  a character (in a somewhat generalised sense).  It represents an
  isometric rotation of $\TSpace{T}{}$ by the angle $a$.
  Under the present conditions we have from~\eqref{eq:def-s-b}:
  \begin{displaymath}
    r(g^{-1}*s(\n{u}))=\matr{ \frac{-\n{b}\n{u}+\n{a}}{ 
        \modulus{-\n{b}\n{u}+\n{a}} } }{0}{0}{ \frac{-\n{b}\n{u}+\n{a}}{ 
        \modulus{-\n{b}\n{u}+\n{a}} } }, \qquad 
    g^{-1}=\matr{\n{a}}{\n{b}}{-\n{b}}{\n{a}}.
  \end{displaymath}
  
  Furthermore we can construct a realization of~\eqref{eq:l-rep} on the 
  functions defined on $ \TSpace{T}{} $ by the formula:
  \begin{equation} \label{eq:ind-b}
    [\rho_\sigma(g) f](\n{v})= 
    \frac{(- \n{v} \n{b} + \bar{\n{a}})^{\sigma}}
    {(- \n{b} \n{v} + {\n{a}})^{1+\sigma}}
    f\left( \frac{\n{a}\n{v}+\n{b}}{-\n{b}\n{v} +\n{a}} \right), \quad 
    g^{-1}=\matr{\n{a}}{\n{b}}{-\n{b}}{\n{a}}.
  \end{equation} It is induced by the character $\chi_{\sigma}$ due to
  formula $-\n{b} \n{v}+\n{a}= (cx+d)\n{p}_1+(bx^{-1}+a)\n{p}_2$,
  where $x=e^t$ and it is a cousin of the principal series
  representation (see~\cite[\S~VI.6, Theorem~8]{Lang85},~\cite[\S~8.2,
  Theorem~2.2]{MTaylor86}).  The
  subspaces of vector valued and even number valued functions are
  invariant under~\eqref{eq:ind-b} and the representation is unitary
  with respect to
  the following inner product:
  \begin{displaymath}
    \scalar{f_1}{f_2}_{ \TSpace{T}{} } = \int_{ \TSpace{T}{} }
    \bar{f}_2(t) f_1(t)\, dt.
  \end{displaymath}
  We select function $f_0(\n{u})\equiv 1$ as our vacuum vector, it is
  a singular one in the same sense as $e^{ix\xi}$ is singular for the
  Fourier transform in Example~\ref{ex:wavelets}\ref{it:H-on-R}. Its
  transformations 
  \begin{equation} \label{eq:coher-b}
    f_g(\n{v})=[\rho_\sigma(g) f_0](\n{v})= \modulus{1+\n{u}^2}^{1/2}
    \frac{(- \n{v} \n{b} + \bar{\n{a}})^{\sigma}}{(- \n{b} \n{v} + 
      {\n{a}})^{1+\sigma}}
  \end{equation}
  and in particular the identity $[\rho_\sigma(g)f_0](\n{v})=
  \bar{\n{a}}^\sigma {\n{a}}^{-1-\sigma} f_0(\n{v})=\n{a}^{-1-2\sigma}
  f_0(\n{v})$ for $g^{-1}=\matr{\n{a}}{0}{0}{\n{a}}$ demonstrates that it is
  a vacuum vector.  Thus we define the reduced wavelet transform accordingly
  to~\eqref{eq:def-s-b} and~\eqref{eq:wave-tr} by the formula:
  \begin{eqnarray}
    [\oper{W}_\sigma f] (\n{u})                        
    & = &  \modulus{1+\n{u}^2}^{1/2} e_{12}\int_{ \TSpace{T}{} }  
    \frac{(-\n{u} e_1 \n{z}  + \n{1})^\sigma \n{z}^{\sigma} }
    {(- e_1 \n{u} + \n{z})^{1+\sigma}} 
    \,d\n{z} \, f(\n{z})  \label{eq:cauchy-b}
  \end{eqnarray} where $\n{z}=e^{e_{12}t}$ is the new monogenic
  variable and $d\n{z}=e_{12}e^{e_{12}t}\,dt$---its differential. The
  integral~\eqref{eq:cauchy-b} is a singular one, its four singular
  points are intersections of the light cone with the origin in
  $\n{u}$ with the unit circle $\TSpace{T}{}$.
  
  The explicit similarity between~\eqref{eq:cauchy} and~\eqref{eq:cauchy-b}
  allows us to consider transformation $\oper{W}_\sigma$~\eqref{eq:cauchy-b} as
  an analog of the Cauchy integral formula and the linear space $
  \FSpace{H}{\sigma}( \TSpace{T}{}) $ generated by the
  coherent states $f_\n{u}(\n{z})$~\eqref{eq:coher-b} as the correspondence
  of the Hardy space $\FSpace{H}{2}(\Space{T}{})$. 
\end{myenumerate}
\end{example}

\subsection{The Dirac (Cauchy-Riemann) and Laplace Operators}
Consideration of Lie groups is hardly possible without consideration of 
their Lie algebras, which are naturally represented by left and right
invariant vectors fields on groups. On a homogeneous space $X=G/H$ we
have also defined a left action of $G$ and can be interested in left
invariant vector fields (first order differential operators). Due to the 
irreducibility of $ \FSpace{F}{2}(x)$ under left action of $G$ every
such vector field $D$ restricted to $ \FSpace{F}{2}(x)$ is a scalar
multiplier of identity $D|_{\FSpace{F}{2}(x)}=cI$. We are
in particular interested in the case $c=0$. 
\begin{defn} \cite{AtiyahSchmid80,KnappWallach76}
  A $G$-invariant first order differential operator 
  \begin{displaymath}
    D_\rho: \FSpace{C}{\infty}(X, \mathcal{S} \otimes V_\rho) 
    \rightarrow
    \FSpace{C}{\infty}(X, \mathcal{S} \otimes V_\rho)
  \end{displaymath} such that $\oper{W}(\FSpace{F}{2}(X))\subset
  \object{ker} D_\rho$ is called \emph{(Cauchy-Riemann-)Dirac
  operator} on $X=G/H$ associated with an irreducible representation $
  \rho $ of $H$ in a space $V_\rho$ and a spinor bundle $\mathcal{S}$.
\end{defn}
The Dirac operator is explicitly defined by the 
formula~\cite[(3.1)]{KnappWallach76}:
\begin{equation} \label{eq:dirac-def}
  D_\rho= \sum_{j=1}^n \rho(Y_j) \otimes c(Y_j) \otimes 1,
\end{equation}
where $Y_j$ is an orthonormal basis of
$\algebra{p}=\algebra{h}^\perp$---the orthogonal completion of the Lie
algebra $\algebra{h}$ of the subgroup $H$ in the Lie algebra $\algebra{g}$
of $G$; $\rho(Y_j)$ is the infinitesimal generator of the right action of
$G$ on $X$; $c(Y_j)$ is Clifford multiplication by $Y_i \in
\algebra{p}$ on the Clifford module $\mathcal{S}$.  We also define 
an invariant Laplacian by the formula
\begin{equation} \label{eq:lap-def}
  \Delta_\rho= \sum_{j=1}^n \rho(Y_j)^2 \otimes \epsilon_j \otimes 1,
\end{equation}
where $\epsilon_j = c(Y_j)^2$ is $+1$ or $-1$.
Note that the Dirac operator~\eqref{eq:dirac-def} is not a factor of
the Laplacian~\eqref{eq:lap-def} unless all commutators $[Y_i,Y_j]$
vanish. Thus null-solutions of $D_\rho$ is not necessarily the
null-solutions of $\Delta_\rho$ \emph{a priory}. But this happens
under our assumptions.
\begin{prop} \textup{\cite{Kisil97c}}
  \label{pr:dirac-laplace}
  Let all commutators of vectors of $ \algebra{h}^\perp $ belong to $
  \algebra{h}$, i.e.
  $[\algebra{h}^\perp,\algebra{h}^\perp]\subset\algebra{h}$.  Let also $f_0$
  be an eigenfunction for all vectors of $ \algebra{h} $ with eigenvalue $0$
  and let also $\oper{W}f_0$ be a null solution to the Dirac operator $D$.
  Then $\Delta f(x)=0$ for all $f(x)\in \FSpace{F}{2}(X)$.
\end{prop}
\begin{example}
  \label{ex:dirac-laplace}
\begin{myenumerate}
\item 
  \label{it:dirac-H-on-R}
  With $G=\Space{H}{}$, $H=\Space{R}{2}$, and $X=\Space{R}{}$ the
  $L_2(X)$ is irreducible therefore both the Dirac and Laplace
  operators are identically zero.
\item 
  \label{it:dirac-H-on-C}
  With $G=\Space{H}{}$, $H=\Space{R}{}$, and $X=\Space{R}{2}$ we
  could take just $\Space{C}{}$ as a ``Clifford algebra'' sufficient
  in that case. The orthogonal completion to the centre $(s,0,0)$
  generates two dimensional Euclidean shifts on
  $\Space{R}{2}=\Space{C}{}$,
  cf. Example~\ref{ex:G-acts-on-X}\ref{it:H-on-C}. As an orthogonal
  basis in that subspace we could take the differential operators
  $\partial/\partial x_1$ and $\partial/\partial x_2$, then we got the
  following realisation of~\eqref{eq:dirac-def} and~\eqref{eq:lap-def}:
  \begin{displaymath}
    D=\frac{\partial}{\partial x_1}-i\frac{\partial}{\partial x_2}
    \qquad \mathrm{ and } \qquad
    \Delta=\frac{\partial^2}{\partial x_1^2}+\frac{\partial^2}{\partial x_2^2},
  \end{displaymath}
  i.e. the classic Cauchy-Riemann and Laplace operators. This is a 
  particular case of invariant operators on nilpotent Lie groups
  considered in~\cite{ConMosc82} and the inclusion
  $[\algebra{h}^\perp,\algebra{h}^\perp]\subset\algebra{h}$ needed in
  Proposition~\ref{pr:dirac-laplace} follows from the nilpotency of
  $\Space{H}{}$. 
\item Let $G=\SL$ and $H$ be its one-dimensional compact subgroup
$K$. Then $\algebra{h}^\perp$ is spanned by two vectors $Y_1=A$ and
$Y_2=B$. In such a situation we can again use $ \Space{C}{} $ instead
of the Clifford algebra $ \Cliff[0]{2} $. Then
formula~\eqref{eq:dirac-def} takes a simple form
$D=\rho(A+iB)$. Infinitesimal action of this operator in the
upper-half plane follows from calculation in~\cite[VI.5(8),
IX.5(3)]{Lang85}, it is $[D_{ \Space{H}{} } f] (z)= -2i y \frac{
\partial f(z)}{ \partial \bar{z} } $, $z=x+iy$. Making the Cayley
transform we can find its action in the unit disk $D_{ \Space{D}{} }
$: again the Cauchy-Riemann operator $ \frac{ \partial }{ \partial
\bar{z} } $ is its principal component.  An explicit calculation of
$D_{ \Space{H}{} }$ was done in~\cite{Kisil97c} and gives the
expected answer
\begin{displaymath}
  D_{ \Space{H}{} }= i\rho(A) + \rho( B) = 2y i\partial_2 + 2y
  \partial_1= 2y \frac{ \partial }{ \partial \bar{z} },
\end{displaymath}
which is just a conformal invariant variant of the
\emph{Cauchy-Riemann equation}. The corresponding operator $\Delta$ is
an invariant Laplacian.

\item
In $\Space{R}{1,1}$  the subgroup 
$H=A$ and its orthogonal completion is spanned by $B$ and $Z$. Thus the
associated Dirac operator has the form $D=e_1 \rho(B) + e_2 \rho(Z)$. 
We need infinitesimal generators of the right action $\rho$ on the ``left'' 
half plane $\TSpace{H}{}$. Following to~\cite{Kisil97c} we find that:
\begin{displaymath}
D_{ \TSpace{H}{} }= e_1\rho(Z) + e_2\rho( A) = 2y (e_1\partial_1 + e_2
\partial_2).
\end{displaymath}
In this case the Dirac operator is not elliptic and as a consequence
we have in particular a singular Cauchy integral
formula~\eqref{eq:cauchy-b}.  Another manifestation of the same
property is that primitives in the ``Taylor expansion'' do not belong
to $\FSpace{F}{2}(\TSpace{T}{})$ itself, see
Example~\ref{ex:taylor}\ref{ex:taylor-b}.  It is known that solutions
of a hyperbolic system (unlike the elliptic one) essentially depend on
the behaviour of the boundary value data.  Thus function theory in
$\Space{R}{1,1}$ is not defined only by the hyperbolic Dirac equation
alone but also by an appropriate boundary condition. The
operator~\eqref{eq:lap-def} in this case is the
\emph{wave  equation} $\Delta= y^{2}(\partial_1^2-\partial_2^2)$.  
\end{myenumerate}
\end{example}

\subsection{The Taylor Expansion}
Both the wavelet transform and its inverse are based on the family
of coherent states $f_a$. For any decomposition 
\begin{equation}\label{eq:cauchy-decomposition}
  f_a(x)=\sum_\alpha \psi_\alpha(x) V_\alpha(a)
\end{equation}
of the coherent states $f_a(x)$ by means of functions $V_\alpha(a)$
(where the sum can become eventually an integral) we have the
\emph{Taylor expansion} 
\begin{eqnarray} 
\hat{f}(a) & = & \int_X f(x) \bar{f}_a(x)\, dx= \int_X f(x) \sum_\alpha 
\bar{\psi}_\alpha(x)\bar{V}_\alpha(a)\, dx  \nonumber \\
 & = &  \sum_\alpha 
\int_X f(x)\bar{\psi}_\alpha(x)\, dx \bar{V}_\alpha(a) 
  =  \sum_{\alpha}^{\infty} \bar{V}_\alpha(a) f_\alpha,\label{eq:taylor}
\end{eqnarray} where $f_\alpha=\int_X f(x)\bar{\psi}_\alpha(x)\, dx$.
However to be useful within the presented scheme such a decomposition
should be connected with the structures of $G$, $H$, and the
representation $\rho$. We will use a decomposition of $f_a(x)$ by the
eigenfunctions $V_\alpha$ of the operators $\rho(h)$, $h\in
\algebra{h}$.
\begin{defn}
 Let $\FSpace{F}{2}(X)=\int_{A} \FSpace{H}{\alpha}\,d\alpha$ be a spectral 
decomposition with respect to the operators $\rho(h)$, $h\in \algebra{h}$.
Then the decomposition
\begin{equation} \label{eq:spec-c}
 f_a(x)= \int_{A} V_\alpha(a) f_\alpha(x)\, d\alpha,
\end{equation} where $f_\alpha(x)\in \FSpace{H}{\alpha}$ and
$V_\alpha(a): \FSpace{H}{\alpha} \rightarrow \FSpace{H}{\alpha}$ is
called the \emph{Taylor decomposition} of the Cauchy kernel $f_a(x)$.
\end{defn}
Note that the Dirac operator $D$ is defined in the terms of left
invariant vector fields and therefore commutes with all
$\rho(h)$. Thus it also has a spectral decomposition over spectral
subspaces of $\rho(h)$:
\begin{equation} \label{eq:spec-d}
 D= \int_{A} D_\delta \, d\delta.
\end{equation}
\begin{prop} \label{pr:cauchy-dirac} \textup{\cite{Kisil97c}}
  The following two equivalent statements links together the Dirac
  operator and the Taylor decomposition in mathematical and physical
  languages respectively.
  \begin{enumerate}
  \item  If spectral measures $d\alpha$ and $d\delta$
    from~\eqref{eq:spec-c} and~\eqref{eq:spec-d} have disjoint
    supports then the image of the Cauchy integral belongs to the
    kernel of the Dirac operator.
  \item We say that $V_a$ is \emph{negative} if $DV_a\neq 0$ and
    $V_a$ is \emph{positive} $\psi_a\neq 0$ in the
    decomposition~\eqref{eq:cauchy-decomposition}. If the intersection
    of positive and negative states is void then the physical states
    from $\FSpace{F}{}(X)$ are null solutions of the Dirac operator.
\end{enumerate}
\end{prop}

\begin{example}
\label{ex:taylor}
\begin{myenumerate}
\item \label{it:taylor-H-on-R}
  For $G=\Space{H}{}$ and $H=\Space{R}{2}$ the only eigenvectors
  of $H$ in the Schr\"odinger representation~\eqref{eq:schrodinger}
  are exponents $e^{ix\xi}$ and the ``Taylor decomposition'' over them
  is in fact the Fourier integral. Note that singularity of the
  vacuum, like in the case~\ref{ex:taylor-b} below, implies that the
  primitive monomials are also outside our space
  $\Space{F}{}(X)$. Here the support of $d\alpha$ is $\Space{R}{}$ and
  from Example~\ref{ex:dirac-laplace}\ref{it:dirac-H-on-R} the support
  of $d\delta$ is the empty set, i.e they are disjoint.
\item For $G=\Space{H}{}$ and $H=\Space{R}{}$ the subgroup $H$ acts
  trivially as multiplication by a scalar on any function thus leave
  us an excessive freedom in th choice of the Taylor decomposition. We
  may wish to use Proposition~\ref{pr:cauchy-dirac} as a guideline. 
  Example~\ref{ex:dirac-laplace}\ref{it:dirac-H-on-C} tells that the
  Dirac operator is the Cauchy-Riemann operator $\partial/\partial
  \bar{z}$. Thus to get a decomposition over a disjoint support we may
  chose the monomial $z^k$ for the Taylor decomposition. The same
  choice is dictated if we wish to obtained the minimum uncertainty
  states~\cite{Perelomov86}.
\item \label{ex:taylor-a}
  Let $G=\SL$ and $H=K$ be its maximal compact subgroup and $\rho_1$
  be described by~\eqref{eq:g-transform}.  $H$ acts on $\Space{T}{}$
  by rotations.  It is one dimensional and eigenfunctions of its
  generator $Z$ are parametrised by odd integers (due to compactness
  of $K$).  Moreover, on the irreducible Hardy space these are
  positive odd integers $n=1,3,5\ldots$ and corresponding
  eigenfunctions are $f_{2n+1}(\phi)=e^{in\phi}$. Negative integers
  span the space of anti-holomorphic function and the splitting
  reflects the existence of analytic structure given by the
  Cauchy-Riemann equation from
  Example~\ref{ex:dirac-laplace}\ref{ex:taylor-a}. The decomposition
  of coherent states $f_a(\phi)$ by means of this functions is well
  known:
  \begin{displaymath}
    f_a(\phi)= \frac{ \sqrt[]{1- \modulus{a}^2 }}{
    \bar{a}e^{i\phi}-1}= \sum_{n=1}^\infty \sqrt[]{1- \modulus{a}^2
    }\bar{a}^{n-1} e^{i(n-1)\phi}= \sum_{n=1}^\infty V_n(a)f_n(\phi),
  \end{displaymath} where $V_n(a)=\sqrt[]{1- \modulus{a}^2
  }\bar{a}^{n-1} $. This is the classical \emph{Taylor expansion} up
  to multipliers coming from the invariant measure.

\begin{table}[pt]
  \begin{center}
   \rotatebox{90}{\begin{minipage}{\textheight}
     \noindent\begin{tabular}{|| c || c | c | c | c | c ||}  
      \hline
      \hline
      \textbf{Notion} & $G=\Space{H}{}$, $H=\Space{R}{2}$ & 
      $G=\Space{H}{}$, $H=\Space{R}{2}$ & 
      $G=\SL$, $H=K$ & $G=\SL$, $H=A$ & $\cdots$ \\[2mm]
      \hline \hline
      Space $X$ & Real line $\Space{R}{}$ & Complex line $\Space{C}{}$
      & Unit disk $\Space{D}{}$  & Disk on
      Fig.~\ref{fig:unit-disk}(b) & $\cdots$ \\[2mm] 
      \hline
      Physical meaning & Configuration space & Phase space 
      & Conformal 2d space   & Space-time (2d) & $\cdots$ \\[2mm] 
      \hline
      Functional space $B$ & $\FSpace{L}{p}(\Space{R}{})$ &
      $\FSpace{L}{p}(\Space{C}{})$ & $\FSpace{L}{p}(\Space{T}{})$&
      $\FSpace{L}{p}(\TSpace{T}{})$ &  $\cdots$ \\[2mm] 
      \hline
      Representation& Schr\"odinger & Schr\"odinger & Conformal maps&
      Conformal maps&  $\cdots$ \\[2mm] 
      \hline
      Vacuum vector& $f_0(x)\equiv 1$& $f_0(x)=e^{-x^2/2}$  &
      $f_0(t)\equiv 1$ & $f_0(t)\equiv 1$ & $\cdots$ \\[2mm] 
      \hline
      Wavelet transform $\oper{W}$ & Fourier transform &
      Segal-Bargmann (SB) tr. & 
      Cauchy integral& Formula~\eqref{eq:cauchy-b} &  $\cdots$ \\[2mm]
      \hline
      Inverse transform $\oper{M}$  & Fourier transform & Inverse SB transform&
      Szeg\"o projection &  &  $\cdots$ \\[2mm]
      \hline
      Image $\FSpace{F}{}(X)$ & Whole $\FSpace{L}{2}(\Space{R}{})$&
      Segal-Bargmann space& Hardy space & Monogenic space &  $\cdots$ \\[2mm]
      \hline
      Dirac operator & Null operator & Cauchy-Riemann oper. & Cauchy-Riemann
      op. & Dirac operator &  $\cdots$ \\[2mm]
      \hline
      Laplace operator & Null operator & Laplace operator & Laplace
      operator & Wave operator &  $\cdots$ \\[2mm]
      \hline
      Taylor expansion & Fourier integral& Taylor expansion & Taylor
      expansion & Formula~\eqref{eq:taylor-b} &  $\cdots$ \\[2mm] 
      \hline
      Image of convolution \cite{Kisil98a}& Weyl
      PDO~\eqref{eq:weyl-calculus} \cite{Folland89,MTaylor81}& Wick
      operator 
      \cite{Berezin88,Folland89}& T\"oplitz operator &
      Sing.Int.Op.~\cite{MTaylor81} &  $\cdots$ \\[2mm]  
      \hline
      \hline
    \end{tabular} \vspace{1cm}                 
    \caption[Periodic table]
    {Periodic table of elements of analytic function theory. The
      dots at the end symbolise an absence of the table end.}
    \label{ta:names}
  \end{minipage}
  }
\end{center}                  
\end{table}  

\item \label{ex:taylor-b}
Let $G=\SL$, $H=A$, and  $\rho_\sigma$ be described by~\eqref{eq:ind-b}.
Subgroup $H$ acts on $\TSpace{T}{}$ by hyperbolic rotations: 
\begin{displaymath}
  \rho: \n{z}=e_1 e^{e_{12}t} \rightarrow e^{2e_{12}\tau}\n{z}=e_1
  e^{e_{12}(2\tau+t)}, \qquad t, \tau \in \TSpace{T}{}.
\end{displaymath}
Then for every $p\in \Space{R}{}$ the function 
$f_p(\n{z})=(\n{z})^p=e^{e_{12}pt}$ where $\n{z}=e^{e_{12}t}$ is an
eigenfunction in the representation~\eqref{eq:ind-b} for a generator $a$ 
of the subgroup $H=A$ with the eigenvalue $2(p-\sigma)-1$,
cf. with~\ref{it:taylor-H-on-R} above.  Again, due to
the analytical structure reflected in the Dirac operator, we only need
negative values of $p$ to decompose the Cauchy integral kernel.
Thereafter for a function $f(\n{z})\in \FSpace{F}{2}(\TSpace{T}{})$ we 
have the following Taylor expansion of its wavelet
transform~\cite{Kisil97c}: 
\begin{equation}\label{eq:taylor-b}
  [\oper{W}_0 f](u)= \int_0^\infty \frac{(e_1\n{u})^{[p]}-1}{e_1\n{u}-1} 
  f_p \,dp,
\qquad \mathrm{where} \quad
  f_p = \int_{\TSpace{T}{}} t \n{z}^{-p}\,d\n{z} f(\n{z}).
\end{equation}
The last integral is similar to the Mellin 
transform~\cite[\S~III.3]{Lang85}, \cite[Chap.~8, (3.12)]{MTaylor86}, 
which naturally arises from the principal series representations 
of $\SL$. 
\end{myenumerate}
\end{example}

Our presentation fits into Table~\ref{ta:names}, which is an extended
version of the table from paper~\cite{Kisil97a}. It was called there
the ``periodic table'' of function theory and it was guessed that it
could help to discover new analytic function theories just like the
periodic table of chemical elements by D.~Mendeleev
help to find new elements. The chain~\ref{it:SL-hyperbolic} of above
examples was presented in~\cite{Kisil97c} as a sample of such a
theory. Not all theretically possible function theory are reasonable
just like not all theoretical combination of protons and neutrons
gives a stable chemical atom. However there are still some more
interesting function theories to descover.

We interrupt here with our overview of the coherent state construction,
but much more could be derived from it. See
books~\cite{AliAntGaz,Perelomov86} for more examples of groups and
physical aspects of the theory. Paper~\cite{Kisil98a} contains further
connections with operator theory, symbolic calculus, PDO, and Toeplitz
operators which we did not mention here at all, see the last raw
of the Table~\ref{ta:names} however.  

\section{Functional Calculus as an Intertwining Operator}
\label{sec:funct-calc-as}

\epigraph{Each problem that I solved became a rule which served
  afterwards to solve other problems.} 
{Descartes}{Discours de la M\'ethode, 1637}\medskip

We saw in the preceeding section that coherent states naturally
generate many principal objects of analysis and function
theory. Particularly the wavelet transform as an intertwining 
operator links different spaces with the same symmetry.
Could we export this observation to other areas? There are many cases
where function spaces provide a model for more complicated objects,
i.e. a functional calculus of operators and quantisation procedure. 
Thus they are the first candidates for an application of that technique.

There are several types of functional calculi, e.g. for selfadjoint
bounded~\cite[\S~VII.2]{SimonReed80} or
unbounded~\cite[\S~VIII.3]{SimonReed80} operators, normal
operators~\cite[\S~VII.3]{DunfordSchwartzI}, several commuting
selfadjoint operators~\cite{JTaylor70}. All of them are defined as
\emph{algebraic homomorphisms} from an algebra of function to an
algebra of operators with some additional properties. On the language
of identification~\eqref{eq:coord-algebra} this means that functional
and operator spaces are forced to have the \emph{same system of
coordinates}. The only exception from that rule is the Weyl functional
calculus~\cite{Anderson69} which is defined just by an integral
formula and is not traditionally obliged to preserve any algebraic
structure. That calculus is a generalisation of the \emph{Weyl
quantisation} \cite[\S~2.1]{Folland89}
\begin{equation}\label{eq:weyl-calculus}
  a_h(p,q) \mapsto a_h(D,X)=\iint \hat{a}(x,y)\sigma_h(0,x,y)\,dx\,dy 
\end{equation}
obtained by an integration of the Schr\"odinger representation $\sigma_h$
~\eqref{eq:schrodinger}. 
 
However there is no need to be so restricted in our choice: we could
give a meaningful general definition~\cite{Kisil95i} of functional
calculus without any references to an algebraic homomorphism property
at all. Moreover we can define a functional calculus not only for an
operator algebra $\algebra{A}$ but also for any $\algebra{A}$-module
$M$~\cite{Kisil02a}.  Let there be two continuous representations
$\rho_f$ and $\rho_a$ of the same topological group $G$ such that
$\rho_f$ act on function space $\FSpace{F}{}(X,\Space{C}{})$ with a
vacuum vector $b_0\in\FSpace{F}{}(X,\Space{C}{})$ and action of
$\rho_a$ on an $M$-valued function space $\FSpace{F}{}(X,M)$ in a way
determined by an element $a\in\algebra{A}$. 

\begin{defn} 
  \label{de:func-calc-sl2} An \emph{analytic functional calculus} of
  an element $a\in \algebra{A}$ is a continuous linear map $\Phi:
  \FSpace{F}{} (X,\Space{C}{}) \rightarrow \FSpace{F}{} (X,
  M):  f(w) \mapsto [\Phi f]$
  if the following conditions fulfil
  \begin{enumerate}
  \item $\Phi$ is an intertwining operator between $\rho_f$ and
    $\rho_a$, namely
    \begin{equation}\label{eq:rho-a-inter}
      [\Phi\rho_{1}(g)f]= \rho_a(g)[\Phi f],
    \end{equation}
    for all $g\in G$ and $f\in \FSpace{F}{}
    (\Space{A}{},\Space{C}{})$. 
  \item $\Phi$ maps the vacuum vector $b_0$ for the
    representation $\rho_f$ to the vacuum vector $b_a\in
    \FSpace{F}{} (X,M)$ for the 
    representation $\rho_a$:
    \begin{equation}\label{eq:rho-a-vac1}
      [\Phi b_0] = b_a.
    \end{equation}
  \end{enumerate}
\end{defn}

We will illustrate advantages of this approach by a construction of
functional calculus for a finite dimensional non-normal matrix. We use
to this end the group $\SL$ together with its discrete series mock
representation $\rho_1$~\eqref{eq:g-transform} in
$\FSpace{L}{2}(\Space{T}{})$ as a model~\cite{Kisil02a} for a
corresponding representation related to a matrix. Let $\algebra{A}$ be
the algebra of complex valued $n\times n$ matrix, $e$ be its unit and
$a\in\algebra{A}$ have all its eigenvalues in the unit circle
$\Space{D}{}$. Then we could define linear-fraction transformation
$g\cdot a$ of $a$ by the formula:
\begin{equation}\label{eq:transform}
  g^{-1}a =(\Ba a - \Bb e)( \alpha e -\beta a )^{-1}, \qquad \mathrm{where}
  \quad g^{-1}=  \matr{ \Ba}{-\Bb}{-\beta}{\alpha},
\end{equation} in analogy with the point
transformation~\eqref{eq:fr-lin-a}.  The set all
matrices~\eqref{eq:transform} for $g\in\SL$ becomes a
$\SL$-homogeneous space. Similarly the \emph{resolvent} $R(g^{-1}a)=(
\Ba e -\Bb a )^{-1}$ is well defined for all $g\in\SL$. Let
$M=\Space{C}{n}$ be a natural left $\algebra{A}$-module and let us
consider a space of $M$-valued functions on $\Space{D}{}$. Analogously
to~\eqref{eq:g-transform} we define a representation $\rho_a$ of $\SL$
as follows:
\begin{equation}\label{eq:rho-a-def}
  (\rho_a(g) f)(z)=R(g^{-1}a)f\left(\frac{\Ba z - \Bb }{\alpha e -\beta z }
  \right), \qquad \mathrm{where}
  \quad g^{-1}=  \matr{ \Ba}{-\Bb}{-\beta}{\alpha}.
\end{equation}

Let $b_0(z)\equiv1$ be the vacuum vector of $\rho_1$ and let
$\FSpace{H}{2}(\Space{D}{},M)$ the minimal $\rho_a$-invariant space of
$M$-valued functions containing all functions $b_x(z)=x\otimes
b_0(z)$, $x\in M$. 
Then the functional calculus in the sense of
Definition~\ref{de:func-calc-sl2} from $\FSpace{H}{2}(\Space{D}{})$ to
$\FSpace{H}{2}(\Space{D}{},M)$ could be
constructed~\cite[Prop.~2.16]{Kisil98a} with the help of 
intertwining properties from  Proposition~\ref{pr:inter1}. Indeed if
we define $\Phi= \oper{M}_{\rho_a}\oper{W}_{\rho_1}$ then: 
\begin{displaymath}
  \Phi \rho_1(g) = \oper{M}_{\rho_a}\oper{W}_{\rho_1} \rho_1 (g)
  = \oper{M}_{\rho_a}  \lambda(g) \oper{W}_{\rho_1}
  =  \rho_a(g)\oper{M}_{\rho_a} \oper{W}_{\rho_1}=  \rho_a(g) \Phi.
\end{displaymath} The explicit integral formula for $\Phi$ coincides with the
integral formula of Dunford-Riesz analytic functional
calculus~\cite[Thm.~VII.1.10]{DunfordSchwartzI}:
\begin{displaymath}[]
  [\Phi f](x,z)=\int_{\Space{T}{}} f(t)(za-te)^{-1}\,dt x.  
\end{displaymath}

Now we can use the constructed functional calculus to get a better
spectral characterisation of $a$ than just the set of its eigenvalues.
The definition used in~\cite{Anderson69} to define the Weyl
spectrum of operator is suitable for the generalisation.
\begin{defn} \textup{\cite{Kisil02a}}
  \label{de:spectr-general}
  A \emph{spectrum} of an operator $a$ is the support of a functional
  calculus $\Phi: f(x) \mapsto f(a)$.
\end{defn} Because now the functional calculus is an intertwining
operator its support is a collection of indecomposable intertwining
operators between $\rho_1$ and $\rho_a$. The entire space
$\FSpace{H}{2}(\Space{D}{},M)$ splits into $\rho_a$-invariant
subspaces $V_{\lambda,k}$ generated by functions $b_x(z)=x\otimes
b_x(z)$, where $x$ is a $k$th root vector for an eigenvalue $\lambda$,
i.e. $(a-\lambda e)^k x=0$ and $(a-\lambda e)^{k-1} x\neq 0$. Such a
minimal $\rho_a$-invariant subspace $V_{\lambda,k}$ up to similarity
is described by the corresponding pair $(\lambda,k)$,
$\lambda\in\Space{D}{}$, $k\in\Space{N}{}$. To get their
classification we need the following notion.

\begin{defn} \textup{\cite[Chap.~4]{Olver95}}
  Two holomorphic functions have $n$th \emph{order contact} in a point
  if their value and their first $n$ derivatives agree at that point,
  in other words their Taylor expansions are the same in first $n+1$
  terms. 

  A point $(z,u^{(n)})=(z,u,u_1,\ldots,u_n)$ of the \emph{jet space}
  $\Space{J}{n}\sim\Space{D}{}\times\Space{C}{n}$ is the equivalence
  class of holomorphic functions having $n$th contact at the point $z$
  with the polynomial:
  \begin{equation}\label{eq:Taylor-polynom}
    p_n(w)=u_n\frac{(w-z)^n}{n!}+\cdots+u_1\frac{(w-z)}{1!}+u.
  \end{equation}
\end{defn}

For a fixed $n$ each holomorphic function
$f:\Space{D}{}\rightarrow\Space{C}{}$ has $n$th \emph{prolongation}
(or \emph{$n$-jet}) $\object[_n]{j}f: \Space{D}{} \rightarrow
\Space{C}{n+1}$: 
\begin{equation}\label{eq:n-jet}
  \object[_n]{j}f(z)=(f(z),f'(z),\ldots,f^{(n)}(z)).
\end{equation}The graph $\Gamma^{(n)}_f$ of $\object[_n]{j}f$ is a
submanifold of $\Space{J}{n}$ which is section of the \emph{jet
bundle} over $\Space{D}{}$ with a fibre $\Space{C}{n+1}$. We also
introduce a notation $J_n$ for the map $
  J_n:f\mapsto\Gamma^{(n)}_f
$ of a holomorphic $f$ to the graph $\Gamma^{(n)}_f$ of its $n$-jet
$\object[_n]{j}f(z)$~\eqref{eq:n-jet}.

One can prolong any map of function $\psi: f(z)\mapsto [\psi f](z)$ to
a map $\psi^{(n)}$ of $n$-jets by the formula
\begin{equation}\label{eq:prolong-def}
  \psi^{(n)} (J_n f) = J_n(\psi f).
\end{equation} For example such a prolongation $\rho_1^{(n)}$ of the
representation $\rho_1$ of the group $\SL$ in
$\FSpace{H}{2}(\Space{D}{})$ (as any other representation of a Lie
group~\cite{Olver95}) will be again a representation of
$\SL$. Equivalently we could say that $J_n$ \emph{intertwines} $\rho_1$ and
$\rho^{(n)}_1$:
\begin{displaymath}
   J_n \rho_1(g)= \rho_1^{(n)}(g) J_n.
\end{displaymath}
Of course, the representation $\rho^{(n)}_1$ is not irreducible: any jet
subspace $\Space{J}{k}$, $0\leq k \leq n$ is
$\rho^{(n)}_1$-invariant subspace of $\Space{J}{n}$. 

Coming back to our representation $\rho_a$~\eqref{eq:rho-a-def} we
could characterise its minimal component as follows.

\begin{prop} \textup{\cite{Kisil02a}}
  Restriction of $\rho_a$ to $V_{\lambda,k}$ is equivalent to the
  extension $\rho_1^{(k)}$ of $\rho_1$ in the $k$th jet space
  $\Space{J}{k}$. Consequently the spectrum of $a$ (defined via the
  functional calculus $\Phi$) consists of exactly $n$ pairs
  $(\lambda_i,k_i)$, $\lambda_i\in\Space{D}{}$, 
  $k_i\in\Space{N}{}$, $1\leq i \leq n$ some of whom could coincide.
\end{prop}

\begin{figure}[tb]
  \begin{center}
      (a)\includegraphics{nccg2.10}\qquad
      (b)\includegraphics{nccg2.11}
 \caption[Three dimensional spectrum]
   {\href{http://maths.leeds.ac.uk/~kisilv/calc1vr.gif}{
   Is spectrum flat (a) or three dimensional (b)---depends
   from viewpoint}.}
    \label{fig:3dspectrum}
  \end{center}
\end{figure}

\begin{example}
  Let $J_k(\lambda)$ denote the Jordan block of the length $k$ for the
  eigenvalue $\lambda$. On the Fig.~\ref{fig:3dspectrum} there are two
  pictures of the spectrum for the matrix
  \begin{displaymath}
    a=J_3\left(\frac{3}{4}e^{i\pi/4}\right)\oplus 
    J_4\left(\frac{2}{3}e^{i5\pi/6}\right) 
    \oplus J_1\left(\frac{2}{5}e^{-i3\pi/4}\right) \oplus 
    J_2\left(\frac{3}{5}e^{-i\pi/3}\right).
  \end{displaymath} Part (a) represents the conventional two-dimensional
  image of the spectrum, i.e. eigenvalues of $a$, and
  \href{http://maths.leeds.ac.uk/~kisilv/calc1vr.gif}{(b) describes
  spectrum $\spec{} a$ arising from the wavelet construction}. The
  first image did not allow to distinguish $a$ from many other
  essentially different matrices, e.g. the diagonal matrix
  \begin{displaymath}
    \diag\left(\frac{3}{4}e^{i\pi/4}, \frac{2}{3}e^{i5\pi/6}, 
    \frac{2}{5} e^{-i3\pi/4},    \frac{3}{5}e^{-i\pi/3}\right).
  \end{displaymath}
  At the same time the Fig.~\ref{fig:3dspectrum}(b)
  completely characterise $a$ up to a similarity. Note that each point of
  $\spec a$ on Fig.~\ref{fig:3dspectrum}(b) corresponds to a particular
  root vector.
\end{example}
The three dimensional spectrum of matrices obeys the spectral mapping
theorem which is a refined version of the classic theorem about
mapping of eigenvalues. 

\begin{thm}[Spectral mapping] \textup{\cite{Kisil02a}}
  Let $\phi: \Space{D}{} \rightarrow \Space{D}{}$ be a holomorphic
  map, let us define $[\phi_* f](z)=f(\phi(z))$ and its prolongation
  $\phi_*^{(n)}$ onto the jet space $\Space{J}{n}$
  by~\eqref{eq:prolong-def}. Its associated action on the pairs
  $(\lambda,k)$ is given by the formula:
  \begin{displaymath}
    \phi_*^{(n)}(\lambda,k)=\left(\phi(\lambda),
      \left[\frac{k}{\deg_\lambda \phi}\right]\right),
  \end{displaymath}
  where $\deg_\lambda \phi$ denotes the degree of zero of the function
  $\phi(z)-\phi(\lambda)$ at the point $z=\lambda$ and $[x]$ denotes
  the integer part of $x$.  Then 
  \begin{displaymath}
    \spec \phi(a) = \phi_*^{(n)} \spec a.
  \end{displaymath}
\end{thm}

The explicit expression for $\phi_*^{(n)}$ which involves derivatives
of $\phi$ upto $n$th order is known, see for
example~\cite[Thm.~6.2.25]{HornJohnson94}, but was not understood before
as form of spectral mapping.

To finish with this topic we will note that our
Definitions~\ref{de:func-calc-sl2} and~\ref{de:spectr-general} are not
restricted to the case $\SL$ only. They are suitable for any group $G$ and
subgroup $H$ from the Table~\ref{ta:names}. For example, Segal-Bargmann
type calculus was outlined in~\cite{Kisil98a} and monogenic calculus
of several noncommuting operators in~\cite{Kisil95i}. This directions
of research still waits a careful exploration.

\section{Quantisation from the Symplectic Invariance}
\label{sec:quant-from-sypl}

\epigraph{Approach your problem from the right end and begin with the
  answer. Then one day, perhaps you will find the final
  question.}{R. van Gulik}{The Hermit Clad in Crane Feather}
\medskip

It is well known that quantum mechanics is full of paradoxes. The more
general is the following: \emph{there are a lot working tools and
  tricks which give numerical predictions for almost all observable
  effects, but the majority of them are mathematically unsound and
  philosophically obscure}. The basic example is the question of
quantisation itself.

It is oftenly said that quantum mechanics is superior to the classical
one but unfortunately we are not able to percept its glory
directly. To obtain a quantum description we have first to describe a
physical system classically and then \emph{quantise} that
description. A popular scheme of quantisation was given by
\person{Dirac}~\cite{Dirac78}.  It says that in order to quantise a
set of classical observables, which are real valued functions on the
phase space, we prescribe a linear map $\hat{}:f\mapsto \hat{f}$ into
selfadjoint operators on a Hilbert space such that for any two
classical observables $f_1$ and $f_2$  
\begin{equation}\label{eq:quantisation-rule}
  \hat{}:f_1 f_2 \mapsto \frac{1}{2}(\hat{f}_1\hat{f}_2+\hat{f}_2\hat{f}_1
  ), \qquad
  \hat{}:\{f_1,f_2\} \mapsto \frac{1}{ i\myhbar}[\hat{f}_1,\hat{f}_2],
\end{equation}
where
$\{\cdot,\cdot\}$ denotes the Poisson brackets in the phase space and
$[\cdot,\cdot]$---the commutator in the operator algebra.

It is also known from various ``no-go'' theorems~\cite[\S~1.1 and
\S~4.4]{Folland89} that those requirements could not be satisfied
beyond the polynomials of degree $\leq 2$.  On the other hand the Weyl
calculus~\eqref{eq:weyl-calculus} gives a good approximation
to~\eqref{eq:quantisation-rule} for a small Planck constant $\hbar$.
Therefore for a physicist interested in numerical predictions of
measurements the quantisation problem is already solved. But
mathematicians are still unhappy with that answer and actively
investigate other quantisation theories, e.g.  geometric
quantisation, deformation quantisation, quantum groups, etc.

The requirements of Dirac~\eqref{eq:quantisation-rule} are similar to
the algebra homomorphism property for the functional calculus: the
first map in~\eqref{eq:quantisation-rule} prescribe the image of a
product for two function and the second is the homomorphism between
two Lie algebra structures. Therefore
mappings~\eqref{eq:quantisation-rule} are silently based on the
identification~\eqref{eq:coord-algebra}. As we saw in the previous
Section one can get a progress in functional calculus if replaces the
algebraic homomorphism condition by the intertwining property in the
spirit of the assumption~\eqref{eq:coord-homogen}. Such a change in a
definition of quantisation is also possible (see below
Definition~\ref{de:quantisation}) and has the following advantages:
\begin{enumerate}
\item It is based on the first physical principles;
\item It has a well defined solution in
  mathematical sense, which naturally turns to be the Weyl
  quantisation. 
\end{enumerate}

\begin{figure}[tbp]
  \begin{center}
      \includegraphics{nccg2.20}
    \caption[Quantum and classic mechanics
      from the Heisenberg group]{
        \href{http://maths.leeds.ac.uk/~kisilv/nccg2.20.gif}{ 
          The appearance of both quantum and
        classic mechanics from the same source. Automorphisms of
        $\Space{H}{}$ generated by symplectic group $Sp(1)$ do not mix
        Schr\"odinger representations $\rho_\hbar$ with different
        $\hbar$ and act by the metaplectic representation inside each
        of them. In the contrast those automorphisms of $\Space{H}{}$
        act transitively on the set of one dimensional representations
        $\rho_{(p,q)}$ joining them into the phase space
        $\Space{R}{2}$}.} 
    \label{fig:p-mechanics}
  \end{center}
\end{figure}

We will outline briefly how to quantise an elementary classical system
with a phase space $\Space{R}{2}$ using an intertwining condition.
There is a natural candidate for a group $G$: this is the group $\SP$
(isomorphic to our permanent companion $\SL$) of linear
\emph{symplectomorphisms} of $\Space{R}{2}$
\begin{equation}\label{eq:symplecto}
  \tau(g):(p,q)\mapsto (ap+bq,cp+dq), \qquad \mathrm{ where } \quad
  g=\matr{a}{b}{c}{d},
\end{equation}
i.e. transformations preserving the \emph{symplectic
form} $(v_1,v_2)=p_1 q_2-p_2 q_1$,
$v_i=(p_i,q_i)~$\cite[\S~41]{Arnold91}. Note that the symplectic form
enters to the expression~\eqref{eq:g-heisenberg} of the multiplication
law on $\Space{H}{}$ making it noncommutative and pops up at the end
of Example~\ref{ex:G-acts-on-X}\ref{it:H-on-C} as expression for
$r(s(z_1)^{-1}*s(z_2))$. It is not surprising therefore that $\SP$
acts as automorphisms of $\Space{H}{}$ as follows:
\begin{equation}\label{eq:Sp-auto}
  \alpha(g): (s,x,y) \mapsto (s, ax+by, cx+dy), \qquad \mathrm{ where
    }
  \quad g=\matr{a}{b}{c}{d}.
\end{equation}
 In fact \emph{$\SP$ is
exactly the subgroup of non-inner automorphisms of $\Space{H}{}$ which
send a Shr\"odinger representation $\rho_\hbar$~\eqref{eq:schrodinger}
with any parameter $\hbar\neq 0$ to an unitary equivalent
representation}~\cite[\S~1.2]{Folland89}.

More precisely for any automorphism $\alpha(g)$, $g\in\SP$ of
$\Space{H}{}$ the composition $\rho_\hbar \circ \alpha(g)$ is again a
representation, which is unitary equivalent to $\rho_\hbar$. Therefore
should exists a unitary operator $U(g)$ in
$\FSpace{L}{2}(\Space{R}{})$ such that $\rho_\hbar \circ \alpha(g)=U(g)
\rho_\hbar U^{-1}(g)$. Then the correspondence $\mu: g \mapsto U(g)$
is a double valued \emph{metaplectic} representation of
$\SP$~\cite[\S~4.2]{Folland89}, \cite[\S~11.3]{MTaylor86} in
$\FSpace{L}{2}(\Space{R}{})$.  We are ready to give our definition of
quantisation.

\begin{defn}
  \label{de:quantisation} Quantisation $\oper{Q}$ is a linear operator
  from $\FSpace{C}{}(\Space{R}{2})$ to
  $B(\FSpace{L}{2}(\Space{R}{}))$, which intertwines two actions of
  the symplectic group $\SP$: by symplectomorphisms $\Space{R}{2}$ in
  classical mechanics and the metaplectic representation $\mu$ in quantum
  mechanics on $\FSpace{L}{2}(\Space{R}{})$:
  \begin{displaymath}
    \oper{Q}\tau(g)=\mu(g)\oper{Q} \qquad \mathrm{for all} \quad
    g\in\SP.
  \end{displaymath}
\end{defn}

The following ``easy-go'' theorem is just an application of several known
results (e.g.~\cite[Thm.~4.28]{Folland89}) about the metaplectic
representation. 
\begin{thm} \textup{\cite{Kisil02b}}
  The Weyl calculus~\eqref{eq:weyl-calculus} is unique (up to
  equivalence) well-defined 
  solution for the quantisation problem~\ref{de:quantisation}.
\end{thm}

It is interesting to note that the quantisation $\oper{Q}$ which
exactly intertwines symplectomorphisms also approximately intertwines
any canonical transformation of $\Space{R}{2}$ modulo smoothing
operators---this is statement of the important Egorov
theorem~\cite[\S~VIII.1]{MTaylor81} from the theory of PDO. 

One can get even more from our study of automorphism of
$\Space{H}{}$. We recall that the Heisenberg group $\Space{H}{}$
besides of the family of Schr\"odinger representations $\sigma_\hbar$
\eqref{eq:schrodinger} with parameter
$\hbar\in\Space{R}{}\setminus\{0\}$ has only in addition the
\emph{family of one dimensional representations} $\rho_{(p,q)}$:
\begin{equation}\label{eq:H-one-dim}
 \rho_{(p,q)}: (s,x,y) \mapsto  e^{i(xp+yq)},\qquad
 \mathrm{ where }\quad (p,q)\in\Space{R}{2}.
\end{equation} That family is usually just mentioned by accurate
authors in the statement of the Stone-von Neumann
theorem~\cite{Folland89,MTaylor86} but almost never used in any way:
what could we expect interesting from commutative representations in
our age of noncommutative geometry? But let us take a second look
assuming that Nature does not create anything without a purpose.

The representation $\alpha$ \eqref{eq:Sp-auto} of $\SP$ could be
lifted to the action $\alpha^*$ on the dual object (the set of all
unitary irreducible representations) $\hHeisen{}$ of $\Space{H}{}$. As
was mentioned before each Schr\"odinger representation
$\sigma_\hbar$~\eqref{eq:schrodinger} is a fixed point of $\alpha^*$.
\href{http://maths.leeds.ac.uk/~kisilv/nccg2.20.gif}{ In the contrast
the whole family of one dimensional representations $\rho_{(p,q)}$ is
just one orbit for $\alpha^*$: it acts transitively on the set $(p,q)$
by symplectomorphisms}~\eqref{eq:symplecto}. Therefore it is natural
to identify the family of representation~\eqref{eq:H-one-dim} with the
\emph{phase space}. This situation is illustrated by
Figure~\ref{fig:p-mechanics}. From the topology on the dual object
$\hHeisen{}$~\cite[\S~7.2.2]{Kirillov94a} the correct place to put the
entire family~\eqref{eq:H-one-dim} is the point $\hbar=0$. Moreover
representations $\sigma_\hbar$, $\hbar\neq 0$ are dense in
$\Space{R}{2}$---this is a form of the \emph{correspondence principle}
between quantum and classical mechanics. Reader may wish to compare
our Figure~\ref{fig:p-mechanics} with Figs.~6 and~7 in
\cite[\S~7.2.2]{Kirillov94a} and corresponding discussion there of
topology on $\hHeisen{}$. It is slightly speculative but we could also
assume that negative values of $\hbar$ on $\hHeisen{}$ correspond to
\emph{anti-particles} because the minus sign of $\hbar$ reverses the
time flow in the Schr\"odinger equation.

How far could that relation between quantum and classic mechanics be
extend? It turns that we can introduce an appropriate notion of
dynamics~\cite{Kisil00a} on $\Space{H}{}$ which produces quantum and
classical dynamics in corresponding
representations~\eqref{eq:schrodinger} and~\eqref{eq:H-one-dim}. That
unified dynamics is based on the following definition.

\begin{defn} \textup{\cite{Kisil00a}}
  \label{de:u-brackets}
  The \emph{$p$-\-mechanical brackets} of two functions $k_1(s,x,y)$,
  $k_2(s,x,y)$ on the Heisenberg group $\Space{H}{}$ are defined as
  follows: 
  \begin{equation}\label{eq:u-brackets}
    \ub{k_1}{k_2}= \anti(k_1*k_2-k_2*k_1),
  \end{equation}
  where $*$ denotes the group convolution on $\Space{H}{}$ of two
  functions and $\anti$ acts as antiderivative with respect of the
  variable $s$. 
\end{defn}

The main property of $p$-mechanical brackets is a common source for
quantum and classical brackets:
\begin{prop} \textup{\cite{Kisil00a}}
  \label{pr:bracket-repr} The images of $p$-\-mechanical
  brackets~\eqref{eq:u-brackets} under infinite dimensional
  representations $\sigma_\hbar$~\eqref{eq:schrodinger}, $\hbar \neq
  0$ and finite dimensional representations
  $\rho_{(q,p)}$~\eqref{eq:H-one-dim} are quantum commutator and
  Poisson brackets of functions $\hat{k}_1$ and $\hat{k}_2$
  respectively:
  \begin{equation}\label{eq:u-bra-repr}
    \rho(\ub{k_1}{k_2})=
    \left\{
    \begin{array}{ll}
      \displaystyle \frac{1}{ i\hbar}[\hat{k}_1,\hat{k}_2]=
      \frac{1}{ i\hbar}(K_1K_2-K_2K_1), & \ 
      \rho=\sigma_\hbar,\ \hbar\neq 0;\vspace{2mm}\\
      \displaystyle \{\hat{k}_1,\hat{k}_2\}= 
      \frac{\partial \hat{k}_1}{\partial q}
      \frac{\partial \hat{k}_2}{\partial p}
      - \frac{\partial \hat{k}_1}{\partial p}
      \frac{\partial \hat{k}_2}{\partial q}, & \ 
      \rho=\rho_{(q,p)}.
    \end{array}
    \right.
  \end{equation}
\end{prop}

We refer to the paper~\cite{Kisil00a} for further details and
discussion about the $p$-mechanical brackets, dynamics generated by
them and its relations to quantum and classical mechanics. It
convinces that \emph{not only quantum mechanics but classical too
appear from representations of the Heisenberg group}.

\section*{Conclusion}
\label{sec:conclusion}
\epigraph{A mathematical idea should not be petrified in a formalised
axiomatic settings, but should be considered instead as flowing as a
river.}{Sylvester}{1878} \medskip

It is accurate to say that the idea of symmetries and invariants was
dominant and the most fruitful in the physics of XX century. There is
no a contradiction however to hope that its systematical use will
reveal many secrets and explain many mysteries of the Nature in the
future as well. That idea is among very few other which are
strong enough to oppose a collapse of science into many unrelated 
parts. Moreover the admiration of symmetries is common to both
cultures (in the sense of \person{C.P.~Snow}) and could be a bridge
between them.

\section*{Acknowledgments}
\label{sec:acknowledgments}

I am grateful to the editors who published my previous
paper~\cite{Kisil97a} on that topic despite of the obvious doubts it
generated.

\bibliographystyle{plain}
\bibliography{abbrevmr,akisil,analyse,acombin,aphilos,aphysics,algebra,arare}

\end{document}